\documentclass[10pt,conference,a4paper]{IEEEtran}

\usepackage[final]{graphicx}
\usepackage[compress]{cite}
\usepackage[T1]{fontenc}
\usepackage{amsmath}
\usepackage{amsfonts}
\usepackage{amssymb, mathrsfs, amsthm}
\usepackage{latexsym}
\usepackage{algorithm}
\usepackage{algorithmic}
\usepackage{subfigure}

\IEEEoverridecommandlockouts

\begin{document}

\title{PAPR Constrained Power Allocation for Iterative Frequency Domain Multiuser SIMO Detector
\vspace{-0.3cm}}

\author{\begin{large}
Valtteri Tervo$^{*+}$, A T\"{o}lli*, J. Karjalainen\dag, Tad Matsumoto$^{*+}$ \end{large} \\
\{wade, atolli, matumoto\}@ee.oulu.fi, juha.pe.karjalainen@gmail.com \\
*Centre for Wireless Communications, University of Oulu\\
P.O. Box 4500, 90014 University of Oulu, Finland.\\
$^{+}$Japan Advanced Institute of Science and Technology\\
1-1 Asahi-Dai, Nomi, Ishikawa, 923-1292 Japan. \\
\dag Samsung Electronics R\&D Institute UK \\
Falcon Business Park, Hali Building, Vaisalantie 4, 01230 Espoo, Finland \vspace{-0.5cm}
\thanks{This work was supported by Finnish Funding Agency for Technology and Innovation (TEKES),
Academy of Finland, Riitta ja Jorma J.\ Takanen Foundation, Finnish Foundation for Technology Promotion, Walter Ahlstr\"{o}m Foundation, Ulla Tuominen foundation and KAUTE-foundation. This work was also in part
supported by the Japanese government funding program, Grant-in-Aid
for Scientific Research (B), No. 23360170.}}

\maketitle

\thispagestyle{plain}
\pagestyle{plain}

\begin{abstract}
Peak to average power ratio (PAPR) constrained power allocation in single carrier multiuser (MU) single-input multiple-output (SIMO) systems with iterative frequency domain (FD) soft cancelation (SC) minimum mean squared error (MMSE) equalization is considered in this paper. To obtain full benefit of the iterative receiver, its convergence properties need to be taken into account also at the transmitter side. In this paper, we extend the existing results on the area of convergence constrained power allocation (CCPA) to consider the instantaneous PAPR at the transmit antenna of each user. In other words, we will introduce a constraint that PAPR cannot exceed a predetermined threshold. By adding the aforementioned constraint into the CCPA optimization framework, the power efficiency of a power amplifier (PA) can be significantly enhanced by enabling it to operate on its linear operation range. Hence, PAPR constraint is especially beneficial for power limited cell-edge users. In this paper, we will derive the instantaneous PAPR constraint as a function of transmit power allocation. Furthermore, successive convex approximation is derived for the PAPR constrained problem. Numerical results show that the proposed method can achieve the objectives described above.

\end{abstract}

\section{Introduction}


Reducing peak to average power ratio (PAPR) in any transmission system is always desirable as it allows use of more efficient and cheaper amplifiers at the transmitter. Recent work on minimizing the PAPR in single carrier frequency division multiple access (FDMA) \cite{Pancaldi-Vitetta-Kalbasi-Al-Dhahir-Uysal-Mheidat-08} transmission can be found in \cite{Slimane-07,Falconer-11,Yuen-Farhang-Boroujeny-12}, where they propose different precoding methods for PAPR reduction. However, these methods do not take into account the transmit power allocation, the channel nor the receiver.

Due to the problems related to inter-symbol-interference (ISI) and multi-user interference (MUI) in single carrier FDMA, efficient low-complexity channel equalization techniques are required. Iterative frequency domain equalization (FDE) technique can achieve a significant performance gain as compared to linear FDE in frequency selective channels. Therefore, it is considered as the most potential candidate to mitigate ISI and MUI \cite{Yuan-Guo-Wang-Ping-08}. However, to exploit the full merit of iterative receiver, the convergence properties of an iterative receiver needs to be taken into account at a transmitter side. This issue has been thoroughly investigated in \cite{Karjalainen-Codreanu-Tolli-Juntti-Matsumoto-11} where the power allocation to different channels is optimized subject to a quality of service (QoS) constraint taking into account the convergence properties of iterative frequncy domain (FD) soft cancelation (SC) minimum mean squared error (MMSE) multiple input multiple output (MIMO) receiver. The convergence properties were examined by using extrinsic information transfer (EXIT) charts \cite{tenBrink-01}. The concept in \cite{Karjalainen-Codreanu-Tolli-Juntti-Matsumoto-11} has been extended for multiuser systems in \cite{Tervo-Tolli-Karjalainen-Matsumoto-12, Tervo-Tolli-Karjalainen-Matsumoto-13}. In this paper, we will introduce a PAPR constraint for the convergence constrained power allocation (CCPA) problem presented in \cite{Tervo-Tolli-Karjalainen-Matsumoto-13}. In other words, we will minimize the total transmit power in a cell with multiple users while guaranteeing the desired QoS in terms of bit error probability (BEP) and keeping the PAPR always below the desired value. This type of power allocation where PAPR is used as a constraint has not yet been published anywhere else. Hence, in this paper we will present our first results on this topic and the development towards the more practical scenarios will be published in the near future.

The main contributions of this paper are summarized as follows: The power of the transmitted waveform is derived as a function of power allocation and quadrature phase shift keying (QPSK) modulated symbol sequence. The instantaneous PAPR constraint is derived and a local convex approximation of the constraint is formulated via change of variable. The constraint is plugged in to a CCPA problem and solved by successive convex approximation (SCA) algorithm.


\section{System Model} \label{sec: system model}

\begin{figure}[tbp!] \vspace{-0.3cm}
\centering
\subfigure[]{
\includegraphics[angle=-90, width=\columnwidth]{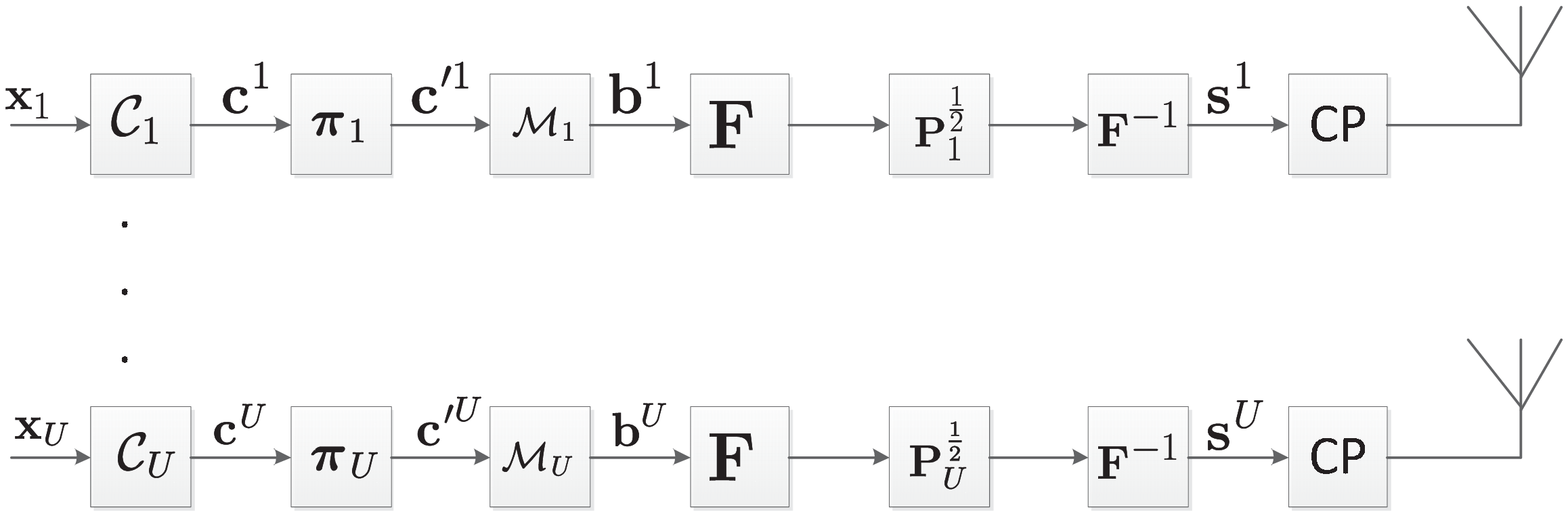}
}
\subfigure[]{
\includegraphics[angle=-90, width=\columnwidth]{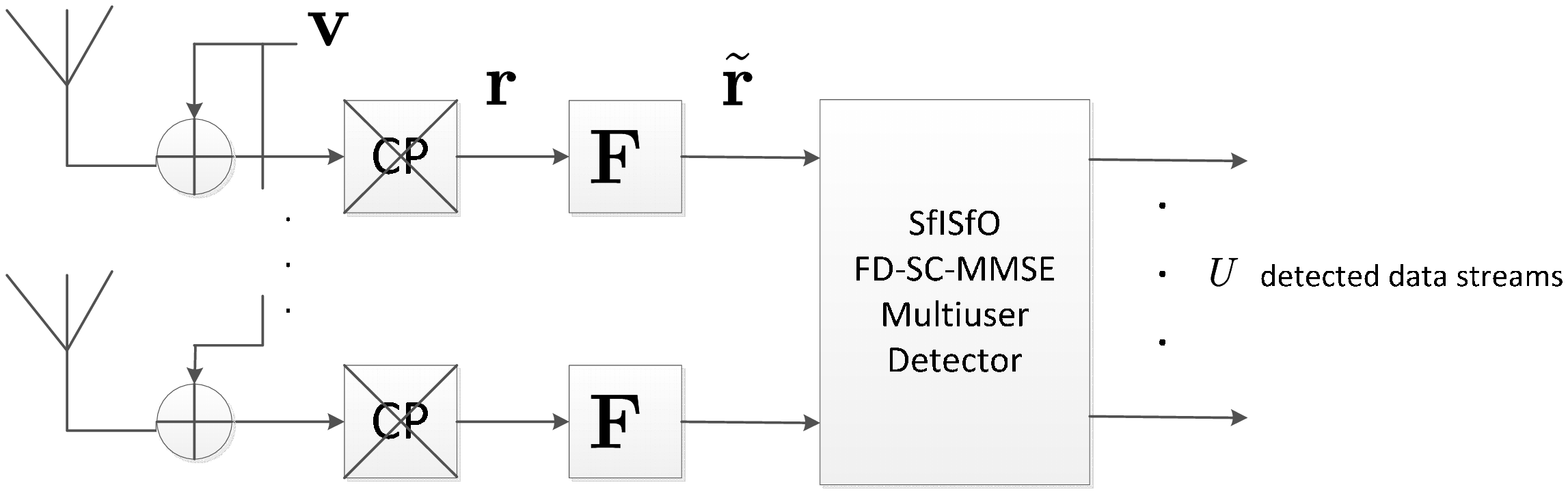}
}
\caption{The block diagram of (a) the transmitter side (b) the receiver side of the system model.}
\label{Fig: system model} \vspace{-0.6cm}
\end{figure}

Consider a single carrier uplink transmission with $U$ single-antenna users and a base station with $N_R$ antennas as depicted in Fig.\ \ref{Fig: system model}. Each user's data stream is encoded by forward error correction code (FEC) $\mathcal{C}_u$, $u=1,2,\dots,U$. The encoded bits are bit interleaved and mapped onto a $2^{N_Q}$-ary complex symbol, where $N_Q$ denotes the number of bits per modulation symbol. After the modulation, each user's data stream is spread across the subcarriers by performing the discrete Fourier transform (DFT) and multiplied with its associated power allocation matrix. Finally, before transmission, each user's data stream is transformed into the time domain by the inverse DFT (IDFT) and a cyclic prefix is added to mitigate inter block interference (IBI).

At the receiver side, after the cyclic prefix removal, the signal can be expressed as
\begin{equation} \label{eq: signal_after_CP_removal}
\mathbf{r}={\bf H}_u{\bf F}^{-1}\mathbf{P}_u^{\frac{1}{2}}\mathbf{F}\mathbf{b}^u+
\sum_{\substack{y=1\\y\ne u}}^{U}{\bf H}_y{\bf F}^{-1}\mathbf{P}_y^{\frac{1}{2}}\mathbf{F}\mathbf{b}^y+\mathbf{v},
\end{equation}
where
${\bf H}_u=[{\bf H}_u^1,{\bf H}_u^2,\dots,{\bf H}_u^{N_R}]^{\text{T}}\in\mathbb{C}^{N_RN_F\times N_F}$ is the space-time channel matrix for user $u$ and
$\mathbf{H}_u^r=\text{circ}\{[h_{u,1}^r,h_{u,2}^r,\dots,h_{u,N_L}^r,\mathbf{0}_{1\times N_F-N_L}]^{\text{T}}\}\in\mathbb{C}^{N_F\times N_F}$ is the time domain circulant channel matrix for user $u$ at the receive antenna $r$. The operator $\text{circ}\{\}$ generates matrix that has a circulant structure of its argument vector and $N_L$ denotes the length of the channel impulse response. ${\bf F}\in\mathbb{C}^{N_F\times N_F}$ denotes the DFT matrix with elements $f_{m,l}=\frac{1}{\sqrt{N_F}}\exp(i2\pi (m-1)(l-1)/N_F)$. $\mathbf{P}\in\mathbb{R}^{UN_F\times UN_F}$ is the power allocation matrix denoted as $\mathbf{P}=\text{diag}(\mathbf{P}_1,\mathbf{P}_2,\dots,\mathbf{P}_{U})$ with ${\bf P}_u=\text{diag}([P_{u,1},P_{u,2},\dots,P_{u,N_F}]^{\text{T}})\in\mathbb{R}^{N_F\times N_F}$, $u=1,2,\dots,U$, and $\mathbf{b}=[{\mathbf{b}^1}^{\text{T}},{\mathbf{b}^2}^{\text{T}},\dots,{\mathbf{b}^{U}}^{\text{T}}]^{\text{T}}$. ${\bf b}^{u}\in\mathbb{C}^{N_F}$, $u=1,2,\dots,U$, is the modulated complex data vector for the $u^{\text{th}}$ user and ${\bf v}\in\mathbb{C}^{N_F}$ is white additive independent identically distributed (i.i.d.) Gaussian noise vector with variance $\sigma^2$.
\section{Problem Formulation} \label{sec: problem formulation}
\begin{figure}[tbp!] \vspace{-0.3cm}
\centering
\includegraphics[angle=-90, width=\columnwidth]{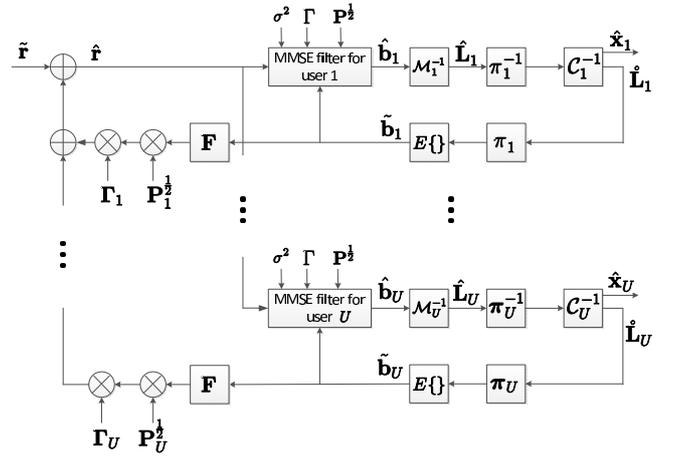}
\caption{The block diagram of FD-SC-MMSE turbo equalizer.}
\label{fig: FD-SC-MMSE} \vspace{-0.6cm}
\end{figure}

In this Section, the characterization of turbo equalizer is given and the derivation of the power minimization problem constrained by the convergence of turbo equalizer is performed.
The block diagram of the FD-SC-MMSE turbo equalizer is depicted in Fig.\ \ref{fig: FD-SC-MMSE}. The frequency domain signal after the soft cancelation can be written as
\begin{equation}
\hat{\mathbf{r}}=\tilde{\mathbf{r}}-\boldsymbol{\Gamma}\mathbf{P}^{\frac{1}{2}}\mathbf{F}_{U}\tilde{\mathbf{b}},
\end{equation}
where $\tilde{\mathbf{b}}=[\tilde{\mathbf{b}^1}^{\text{T}},\tilde{\mathbf{b}^2}^{\text{T}},\dots,
\tilde{\mathbf{b}^{U}}^{\text{T}}]^{\text{T}}\in\mathbb{C}^{UN_F}$ are the soft symbol estimates of the modulated complex symbols and $\mathbf{F}_{U}=\mathbf{I}_{U}\otimes\mathbf{F}\in\mathbb{C}^{UN_F\times UN_F}$. $\mathbf{I}_{U}$ denotes the $U\times U$ identity matrix and $\otimes$ is the Kronecker product. $\boldsymbol{\Gamma}=[\boldsymbol{\Gamma}_1,\boldsymbol{\Gamma}_2,\dots,\boldsymbol{\Gamma}_{U}]
\in\mathbb{C}^{N_RN_F\times UN_F}$ and $\boldsymbol{\Gamma}_u=\text{bdiag}\{\boldsymbol{\Gamma}_{u,1},\boldsymbol{\Gamma}_{u,2},\dots,
\boldsymbol{\Gamma}_{u,N_F}\}\in\mathbb{C}^{N_RN_F\times N_F}$ is the space-frequency channel matrix for user $u$ expressed as
\begin{equation}
\boldsymbol{\Gamma}_u=\mathbf{F}_{N_R}{\bf H}_u\mathbf{F}^{-1}.
\end{equation}
$\boldsymbol{\Gamma}_{u,m}\in\mathbb{C}^{N_R\times N_R}$ is the diagonal channel matrix for $m^{\text{th}}$ frequency bin of $u^{\text{th}}$ user and $\text{bdiag}\{\cdot\}$ generates block diagonal matrix of its arguments.
$\hat{\bf L}_u$ and $\mathring{\bf L}_u$ in Fig.\ \ref{fig: FD-SC-MMSE} denote the log-likelihood ratios (LLRs) provided by the equalizer and the channel decoder of user $u$, respectively, and $\hat{\bf x}_u$ denotes the estimate of ${\bf x}_u$.
The problem formulation follows that presented in \cite{Karjalainen-Codreanu-Tolli-Juntti-Matsumoto-11,Tervo-Tolli-Karjalainen-Matsumoto-12,
Tervo-Tolli-Karjalainen-Matsumoto-13}.
Let $\hat{I}_u^{\text{E}}$ denote the mutual information (MI) between the transmitted interleaved coded bits ${\bf c}'^u$ and the LLRs at the output of the equalizer $\hat{\bf L}_u$. Moreover, let $\hat{I}_u^{\text{A}}$ denote the \textit{a priori} MI at the input of the equalizer and $\hat{f}_u()$ denote a monotonically increasing EXIT function of the equalizer of the $u$th user. Now, we can write the following relationship:
\begin{equation}
\hat{I}_u^{\text{E}}=\hat{f}(\hat{I}_1^{\text{A}},\hat{I}_2^{\text{A}},\dots,\hat{I}_{U}^{\text{A}}).
\end{equation}
Similarly, let $\mathring{I}_u^{\text{E}}$ denote the extrinsic MI at the output of the decoder and $\mathring{I}_u^{\text{A}}$ \textit{a priori} MI at the input of the decoder. We can write
\begin{equation}
\mathring{I}_u^{\text{E}}=\mathring{f}_u(\mathring{I}_u^{\text{A}}),
\end{equation}
where $\mathring{f}_u()$ is a monotonically increasing and, hence, invertible EXIT function of the decoder.

Because interleaving has no impact on the MI, i.e., $\hat{I}_u^{\text{E}} = \mathring{I}_u^{\text{A}}$ and $\hat{I}_u^{\text{A}} = \mathring{I}_u^{\text{E}}$, we can express the condition for keeping the convergence tunnel open for each user as
\begin{align} \label{eq: constr1}
\exists\{\mathring{I}_i^{\text{E}}\in[0,1]\}_{\substack{i=1 \\ i\ne u}}^U:
\hat{f}_u(\mathring{I}_1^{\text{E}},\dots,\mathring{I}_u^{\text{E}},\dots,\mathring{I}_{U}^{\text{E}})
\ge\mathring{f}_u^{-1}
(\mathring{I}_u^{\text{E}})+\epsilon_u \nonumber \\
\forall u=1,2\dots,U,
\end{align}
i.e., for all $u$, there exists a set of outputs from the decoders of all the users except $u$ such that the EXIT function of the equalizer of user $u$ is above the inverse of the EXIT function of the decoder of user $u$ plus a parameter $\epsilon_u$. In other words, the convergence is guaranteed as long as there exists an open tunnel between the decoder and equalizer EXIT surfaces until the convergence point. $\epsilon_u$ is a parameter controlling the minimum gap between the EXIT surfaces.
To make the problem tractable, continuous convergence condition \eqref{eq: constr1} is discretized (see \cite[Section IV]{Karjalainen-Codreanu-Tolli-Juntti-Matsumoto-11} for more details) and replaced with
\begin{align} \label{eq: MIconstr}
\exists\Big\{\mathring{I}_{i,k_i}^{\text{E}}\in[0,1]:k_i\in\{1,2,\dots,K\}\Big\}_{\substack{i=1 \\ i\ne u}}^U: \nonumber \\
\hat{f}_{u}(\mathring{I}_{1,k_1}^{\text{E}},\dots,\mathring{I}_{u,k_u}^{\text{E}},
\dots,\mathring{I}_{U,k_{U}}^{\text{E}})
\ge\mathring{f}_u^{-1}
(\mathring{I}_{u,k_u}^{\text{E}})+\epsilon_{u,k_u}, \nonumber \\
\forall k_u=1,2,\dots,K, \forall u=1,2\dots,U,
\end{align}
where $k_u$ denotes the index of MI point such that $\mathring{I}_{u,k_u+1}^{\text{E}}>\mathring{I}_{u,k_u}^{\text{E}}$, $\forall k_u=1,2,\dots,K-1$, i.e., the indexing is ordered such that the MI increases with the index. In this paper, we assume $\epsilon_{u,k_u}=\epsilon_u$, $\forall k_u<K$ and $\epsilon_{u,K}=0$.

Using the inverse of the J-function \cite{Brannstrom-Rasmussen-Grant-05}\footnote{J-function assumes that the LLRs are Gaussian distributed with variance being equal to two times mean.}, the constraints can be written as
\begin{align} \label{eq: Jconstr}
\exists\Big\{\mathring{I}_{i,k_i}^{\text{E}}\in[0,1]:k_i\in\{1,2,\dots,K\}\Big\}_{\substack{i=1 \\ i\ne u}}^U: \nonumber \\
\text{J}^{-1}(\hat{f}_{u}(\mathring{I}_{1,k_1}^{\text{E}},\dots,\mathring{I}_{u,k_u}^{\text{E}},\dots,
\mathring{I}_{U,k_U}^{\text{E}}))
\ge\text{J}^{-1}(\mathring{f}_u^{-1}
(\mathring{I}_{u,k_u}^{\text{E}})+\epsilon_{u}),  \nonumber \\
\forall k_u=1,2,\dots,K, \forall u=1,2\dots,U.
\end{align}
We will use the so called diagonal sampling \cite{Tervo-Tolli-Karjalainen-Matsumoto-13}, i.e., we choose only the points in the $U+1$-dimensional EXIT space where all the decoder's outputs are equal, i.e., we check the $K$ points on the line from the origin to the convergence point. Although this method is suboptimal, a sophisticated guess is that the active constraints lie on this line due to the smoothness of the decoder surface.
The convergence constraint simplifies to
\begin{align} \label{eq: variance_constraint}
\hat{\sigma}^2_{u,k}\ge\mathring{\sigma}^2_{u,k},
\forall u=1,2\dots,U, \forall k=1,2,\dots,K.
\end{align}
When Gray-mapped quadrature phase shift keying (QPSK) modulation is used, the variance of the LLRs at the output of the equalizer can be expressed as \cite[Eq.\ (17)]{Karjalainen-Codreanu-Tolli-Juntti-Matsumoto-11}
\begin{align} \label{eq: LLRvariance}
\hat{\sigma}^2_{u,k}=\frac{4\zeta_{u,k}}{1-\zeta_{u,k}\bar{\Delta}_{u,k}},
\end{align}
where $\zeta_{u,k}$ is the effective signal-to-interference-plus-noise power ratio (SINR) for $u^{\text{th}}$ user at $k^{\text{th}}$ MI index. Plugging \eqref{eq: LLRvariance} into \eqref{eq: variance_constraint}, the convergence constraint power minimization problem can be expressed as
\begin{equation}
\begin{array}{lll} \label{eq: SIMO_opt_original}
\underset{P_{u,n}, \boldsymbol{\omega}_{u,n}}{\text{minimize}} & \text{tr}\{\mathbf{P}\} &  \\
\text{subject to} & \frac{1}{N_F}\sum_{m=1}^{N_F}\frac{P_{u,m}|{\boldsymbol{\omega}}_{u,m}^{\text{H}}{\boldsymbol\gamma}_{u,m}|^2}
{\sum_{l=1}^{U}P_{l,m}|{\boldsymbol{\omega}}_{u,m}^{\text{H}}{\boldsymbol\gamma}_{l,m}|^2\bar{\Delta}_l+
||\boldsymbol{\omega}_{u,m}||^2\sigma^2} \\ & \ge\xi_{u,k},
\forall u=1,2\dots,U, \forall k=1,2,\dots,K, \\
& P_{u,n}\ge0, \\
& u=1,2,\dots,U, n=1,2,\dots,N_F,
\end{array}
\end{equation}
where
\begin{equation}
\xi_{u,k}=\frac{(\mathring{\sigma}_{u,k})^2}{4+(\mathring{\sigma}_{u,k})^2\bar{\Delta}_{u,k}}
\end{equation}
is constant.
${\boldsymbol\gamma}_{u,m}\in\mathbb{C}^{N_R}$ consists of the diagonal elements of ${\boldsymbol\Gamma}_{u,m}$, i.e., ${\boldsymbol\gamma}_{u,m}$ is the channel vector for $m^{\text{th}}$ frequency bin of user $u$.
${\boldsymbol{\omega}}_{u,m}\in\mathbb{C}^{N_R}$ is the receive beamforming vector for $m^{\text{th}}$ frequency bin of user $u$ and it can be optimally calculated as \cite{Karjalainen-11}
\begin{equation}
{\boldsymbol{\omega}}_{u,m}=\frac{(\sum_{l=1}^{U}P_{l,m}
{\boldsymbol\gamma}_{l,m}
{\boldsymbol\gamma}_{l,m}^{\text{H}}\bar{\Delta}_l+\sigma^2{\bf I}_{N_R})^{-1}{\boldsymbol\gamma}_{u,m}P_{u,m}^{\frac{1}{2}}}{\text{avg}\{\ddot{\bf b}^u\}\zeta_{u,k}+1}.
\end{equation}
$\bar{\Delta}_{u,k}=\text{avg}\{\mathbf{1}_{N_F}-\ddot{\mathbf{b}}^u\}\in\mathbb{R}$ is the average residual interference of the soft symbol estimates and $\ddot{\mathbf{b}}^u=[|\tilde{b}_1^u|^2,|\tilde{b}_2^u|^2,\dots,|\tilde{b}_{N_F}^u|^2]^{\text{T}}\in\mathbb{C}^{N_F}$.
The soft symbol estimate $\tilde{b}_n^u$ is calculated as
\begin{equation}
\tilde{b}_n^u=E\{b_n^u\}=\sum_{b_i\in\mathfrak{B}}b_i\Pr(b_n^u=b_i),
\end{equation}
where $\mathfrak{B}$ is the modulation symbol alphabet, and the symbol \textit{a priori} probability can be calculated by
\begin{align}
\Pr(b_n^u=b_i)&=\prod_{q=1}^{N_Q}\Pr(c_{n,q}^u=z_{i,q}) \nonumber \\
&=\Big(\frac{1}{2}\Big)^{N_Q}\prod_{q=1}^{N_Q}(1-\bar{z}_{i,q}\tanh(\lambda_{n,q}^u/2)),
\end{align}
with $\bar{z}_{i,q}=2z_{i,q}-1$ and ${\bf z}_i=[z_{i,1}, z_{i,2}, \dots, z_{i,N_Q}]^{\text{T}}$ is the binary representation of the symbol $b_i$, depending on the modulation mapping. $\lambda_{n,q}^u$ is the \textit{a priori} LLR of the bit $c_{n,q}^u$, provided
by the decoder of user $u$.

\subsection{Successive Convex Approximation via Variable Change} \label{sec: SCAVC}
In this Section, we derive a successive convex approximation for the non-convex power minimization problem \eqref{eq: SIMO_opt_original}.
Let $\alpha_{u,m}\in\mathbb{R}$, such that $P_{u,m}=e^{\alpha_{u,m}},\forall u=1,2,\dots,U,m=1,2,\dots,N_F$ and $t_{u,m}^k\in\mathbb{R}^+, \forall u=1,2,\dots,U,m=1,2,\dots,N_F, k=1,2,\dots,K$. Since the active inequality constraints in \eqref{eq: SIMO_opt_original} hold with equality at the optimal point, we can express \eqref{eq: SIMO_opt_original} for fixed receive beamformers as
\begin{equation}
\begin{array}{lll} \label{eq: SIMO_opt_equivalent}
\underset{{\boldsymbol \alpha},{\bf t}}{\text{minimize}} & \sum_{u=1}^{U}\sum_{m=1}^{N_F}e^{\alpha_{u,m}} &  \\
\text{subject to} & \frac{1}{N_F}\sum_{m=1}^{N_F}t_{u,m}^k\ge\xi_{u,k} \\
& u=1,2,\dots,U, k=1,2,\dots,K,  \\
(**) & \frac{e^{\alpha_{u,n}}|{\boldsymbol{\omega}_{u,n}^k}^{\text{H}}{\boldsymbol\gamma}_{u,n}|^2}
{
\sum_{l=1}^{U}e^{\alpha_{l,n}}
|{\boldsymbol{\omega}_{u,n}^k}^{\text{H}}{\boldsymbol\gamma}_{l,n}|^2\bar{\Delta}_{k}+
\sigma^2||{\boldsymbol{\omega}_{u,n}^k}||^2}\ge t_{u,n}^k, \\
& k=1,2,\dots,K, u=1,2,\dots,U, \\
& n=1,2,\dots,N_F,
\end{array}
\end{equation}
where the optimization variables are ${\bf t}=\{t_{u,m}^k: u=1,2,\dots,U,k=1,2,\dots,K,m=1,2,\dots,N_F\}$, and ${\boldsymbol \alpha}=\{\alpha_{u,m}: u=1,2,\dots,U,m=1,2,\dots,N_F\}$. By taking the natural logarithm of the constraint $(**)$ yields
\begin{align} \label{eq: logconstr}
&\alpha_{u,n}+2\ln(|{\boldsymbol{\omega}_{u,n}^k}^{\text{H}}{\boldsymbol\gamma}_{u,n}|) \nonumber \\
&-\ln(\sum_{l=1}^{U}e^{\alpha_{l,n}}
|{\boldsymbol{\omega}_{u,n}^k}^{\text{H}}{\boldsymbol\gamma}_{l,n}|^2\bar{\Delta}_{k}+
\sigma^2||{\boldsymbol{\omega}_{u,n}^k}||^2)\ge\ln t_{u,n}^k.
\end{align}
Since a logarithm of the summation of the exponentials is convex, the left hand side (LHS) of the constraint \eqref{eq: logconstr} is concave. The RHS of \eqref{eq: logconstr} can be locally approximated with its best convex upper bound, i.e., linear approximation of $\ln t_{u,n}^k$ at a point $\hat{t}_{u,n}^k$:
\begin{equation} \label{eq: lin.approx.}
Y(t_{u,n}^k,\hat{t}_{u,n}^k)=\ln\hat{t}_{u,n}^k+\frac{(t_{u,n}^k-\hat{t}_{u,n}^k)}
{\hat{t}_{u,n}^k}.
\end{equation}
A local convex approximation of \eqref{eq: SIMO_opt_equivalent} can be written as
\begin{equation}
\begin{array}{lll} \label{eq: local_conv_appr}
\underset{{\boldsymbol \alpha}, {\bf t}}{\text{minimize}} & \sum_{u=1}^{U}\sum_{m=1}^{N_F}e^{\alpha_{u,m}}   \\
\text{subject to} & \sum_{m=1}^{N_F}t_{u,m}^k\ge N_F\xi_{u,k}, u=1,2,\dots,U, \\
& k=1,2,\dots,K, \\
& \alpha_{u,n}+2\ln(|{\boldsymbol{\omega}_{u,n}^k}^{\text{H}}{\boldsymbol\gamma}_{u,n}|)- \\ &\ln(\sum_{l=1}^{U}e^{\alpha_{l,n}}
|{\boldsymbol{\omega}_{u,n}^k}^{\text{H}}{\boldsymbol\gamma}_{l,n}|^2\bar{\Delta}_{k}+
\sigma^2||{\boldsymbol{\omega}_{u,n}^k}||^2)\ge \\ & Y(t_{u,n}^k,\hat{t}_{u,n}^k), u=1,2,\dots,U, \\
& k=1,2,\dots,K, n=1,2,\dots,N_F,
\end{array}
\end{equation}
and it can be solved efficiently by using standard optimization tools, e.g., interior-point methods \cite{Boyd-Vandenberghe-04}.

The SCA algorithm starts by a feasible initialization $\hat{t}_{u,n}^k=\hat{t}_{u,n}^{k(0)}, \forall u,k,n$. After this, \eqref{eq: local_conv_appr} is solved yielding a solution ${t}_{u,n}^{k(*)}$ which is used as a new point for the linear approximation. The procedure is repeated until convergence. \textbf{Algorithm} \ref{alg: SCA} provides the algorithm description for the SCA algorithm.
\vspace{-0.2cm}
\begin{algorithm}
\caption{Successive convex approximation algorithm.}
\label{alg: SCA}
\begin{minipage}{\columnwidth}
\begin{algorithmic}[1]
\STATE Set $\hat{t}_{u,n}^k=\hat{t}_{u,n}^{k(0)}, \forall u,k,n$.
\REPEAT
\STATE Solve Eq.\ \eqref{eq: local_conv_appr}.
\STATE Update $\hat{t}_{u,n}^k={t}_{u,n}^{k(*)}, \forall u,k,n$.
\UNTIL Convergence.
\end{algorithmic}
\end{minipage}
\end{algorithm}
\vspace{-1cm}
\section{Instantaneous PAPR Constraint} \label{sec: Instant_PAPR}
\vspace{-0.1cm}
In this Section, the PAPR constraint is derived.
Because the PAPR is derived similarly for all the users, the user index is omitted in this section. Let ${\bf G}={\bf F}^{-1}{\bf P}^{\frac{1}{2}}{\bf F}$. The entry $(m,n)$ of ${\bf G}$ is obtained as \vspace{-0.3cm}
\begin{equation}
g_{m,n}=\frac{1}{N_F}\sum_{l=1}^{N_F}\sqrt{P_l}e^{\frac{j2\pi(l-1)(n-m)}{N_F}}.
\end{equation}
Let $s_m$ be the $m^{\text{th}}$ output of the transmitted waveform after the IFFT. PAPR can be calculated as
\begin{equation}
\text{PAPR}=\frac{\max_{m}|s_m|^2}{\text{avg}[|s_m|^2]},
\end{equation}
where $s_m=\sum_{n=1}^{N_F}g_{m,n}b_n$.

Assuming $|b_n|=1$, $\forall n$ and $\mathbb{E}\{b_pb_q^*\}=0$, $\forall p\ne q$, where $b_q^*$ denotes the complex conjugate of $b_q$, the average can be calculated as
\begin{align}
\text{avg}[|s_m|^2]=\frac{1}{N_F}\sum_{m=1}^{N_F}\mathbb{E}\Big\{[|s_m|^2]\Big\}
=\frac{1}{N_F}\sum_{l=1}^{N_F}P_l.
\end{align}

The power of the $m^{\text{th}}$ transmitted waveform can be calculated as
\begin{align}
|s_m|^2=&\frac{1}{N_F}\sum_{l=1}^{N_F}P_l+\frac{1}{N_F^2}\sum_{\substack{q,p=1\\p\ne q}}^{N_F}b_pb_q^*\sum_{l=1}^{N_F}P_la_{lpq}+ \nonumber \\
&\frac{1}{N_F^2}\sum_{\substack{q,p=1\\p\ne q}}^{N_F}b_pb_q^*\sum_{\substack{n,i=1\\i\ne n}}^{N_F}\sqrt{P_nP_i}e^{\frac{j2\pi((n-1)(p-m)-(i-1)(q-m))}{N_F}},
\end{align}
where $a_{lpq}=e^{\frac{j2\pi(l-1)(p-q)}{N_F}}$.
This can be simplified to
\begin{align} \label{eq: simplified peak power}
|s_m|^2=\frac{1}{N_F}\sum_{l=1}^{N_F}(1+\frac{2d_l}{N_F})P_l+\frac{2}{N_F^2}\sum_{\substack{n,i=1\\i> n}}^{N_F}
\eta_{nim}\sqrt{P_nP_i},
\end{align}
where
\begin{align}
d_l=&\sum_{\substack{q,p=1\\p>q}}^{N_F}\Big(\mathcal{R}[a_{lpq}](\mathcal{R}[b_p]\mathcal{R}[b_q]+ \nonumber \\
& \mathcal{I}[b_p]\mathcal{I}[b_q])+\mathcal{I}[a_{lpq}](\mathcal{R}[b_p]\mathcal{I}[b_q]-\mathcal{I}[b_p]
\mathcal{R}[b_q])\Big),
\end{align}
and \vspace{-0.3cm}
\begin{align}
\eta_{nim}=&\sum_{\substack{q,p=1\\p>q}}^{N_F}\Big((\mathcal{R}[b_p]\mathcal{R}[b_q]+\mathcal{I}[b_p]\mathcal{I}[b_q])
(\mathcal{R}[a_{npm}a_{iqm}^*]+ \nonumber \\ &\mathcal{R}[a_{nqm}a_{ipm}^*])-(\mathcal{I}[b_p]\mathcal{R}[b_q]-\mathcal{R}[b_p]
\mathcal{I}[b_q])(\mathcal{I}[a_{npm}a_{iqm}^*] \nonumber \\
&-\mathcal{I}[a_{nqm}a_{ipm}^*])\Big).
\end{align} \vspace{-0.1cm}
Operators $\mathcal{R}$ and $\mathcal{I}$ take the real and imaginary part of a complex argument, respectively.

\subsection{Successive Convex Approximation via Variable Change} \label{sec: SCAVCPAPR}
In this Section, we derive a successive convex approximation for the non-convex PAPR constraint.
Due to the nonnegativity of the absolute value, the factor $1+\frac{2d_l}{N_F}$ in \eqref{eq: simplified peak power} has to be nonnegative. However, the factor $\eta_{nim}$ can be negative, depending on the symbol sequence and the power allocation. Let $\eta_{nim}^+=\max\{0,\eta_{nim}\}$ and $\eta_{nim}^-=\min\{\eta_{nim},0\}$.
The instantaneous PAPR constraint can be written as \vspace{-0.25cm}
\begin{align} \label{eq: PAPR constraint}
\sum_{l=1}^{N_F}(1+\frac{2d_l}{N_F})P_l+\frac{2}{N_F}\sum_{\substack{n,i=1\\i>n}}^{N_F}
\eta_{nim}^+\sqrt{P_nP_i}\nonumber \\
\le
\delta\sum_{l=1}^{N_F}P_l-
\frac{2}{N_F}\sum_{\substack{n,i=1\\i>n}}^{N_F}
\eta_{nim}^-\sqrt{P_nP_i}, \forall m=1,2,\dots,N_F,
\end{align}
where $\delta$ is a user specific parameter controlling the PAPR.

Denoting $P_l=e^{\alpha_l}$, $l=1,2,\dots,N_F$, and taking the logarithm from both sides of \eqref{eq: PAPR constraint}, the constraint becomes \vspace{-0.15cm}
\begin{align} \label{eq: SCAVC constaint}
\ln\Big(\sum_{l=1}^{N_F}(1+\frac{2d_l}{N_F})e^{\alpha_l}+\frac{2}{N_F}\sum_{\substack{n,i=1\\i>n}}^{N_F}
\eta_{nim}^+e^{\frac{1}{2}(\alpha_n+\alpha_i)}\Big)\nonumber \\
\le
\ln\Big(\delta\sum_{l=1}^{N_F}e^{\alpha_l}+
\frac{2}{N_F}\sum_{\substack{n,i=1\\i>n}}^{N_F}
(-\eta_{nim}^-)e^{\frac{1}{2}(\alpha_n+\alpha_i)}\Big).
\end{align}
Both sides of \eqref{eq: SCAVC constaint} are convex functions. RHS can be approximated by a linear function and then using the SCA technique similarly to \eqref{eq: lin.approx.} and \eqref{eq: local_conv_appr}, a local solution can be found. Let
\begin{align}
W_m(\boldsymbol{\alpha})= \ln\Big(\delta\sum_{l=1}^{N_F}e^{\alpha_l}+
\frac{2}{N_F}\sum_{\substack{n,i=1\\i>n}}^{N_F}
(-\eta_{nim}^-)e^{\frac{1}{2}(\alpha_n+\alpha_i)}\Big) \nonumber.
\end{align}
The best concave approximation of $W_m(\boldsymbol{\alpha})$ at a point $\hat{\boldsymbol{\alpha}}$ is given by
\begin{equation}
T_m(\boldsymbol{\alpha},\hat{\boldsymbol{\alpha}})=W_m(\hat{\boldsymbol{\alpha}})+\sum_{k=1}^{N_F}\frac{\partial W_m}{\partial\alpha_k}(\hat{\boldsymbol{\alpha}})(\alpha_k-\hat{\alpha}_k).
\end{equation}
The partial derivative $\frac{\partial W_m}{\partial\alpha_k}$ is given by \eqref{eq: PAPR linear approx}.
\begin{figure*}[tbp!] \vspace{-0.3cm}
\small
\begin{eqnarray} \label{eq: PAPR linear approx}
\frac{\partial W_m}{\partial\alpha_k}=\frac{\delta e^{\alpha_k}-\frac{1}{N_F}\sum_{i=k+1}^{N_F}\eta_{kim}^-e^{\frac{1}{2}(\alpha_k+\alpha_i)}-\frac{1}{N_F}
\sum_{n=1}^{k-1}\eta_{nkm}^-e^{\frac{1}{2}(\alpha_n+\alpha_k)}}{\delta\sum_{l=1}^{N_F}e^{\alpha_l}+
\frac{2}{N_F}\sum_{\substack{n,i=1\\i>n}}^{N_F}
(-\eta_{nim}^-)e^{\frac{1}{2}(\alpha_n+\alpha_i)}}.
\end{eqnarray}
\hrulefill \vspace{-0.2cm}
\end{figure*}

The approximation of the PAPR constrained problem is now written as
\begin{equation}
\begin{array}{lll} \label{eq: local_conv_appr2}
\underset{{\boldsymbol \alpha}, {\bf t}}{\text{minimize}} & \sum_{u=1}^{U}\sum_{m=1}^{N_F}e^{\alpha_{u,m}}   \\
\text{subject to} & \sum_{m=1}^{N_F}t_{u,m}^k\ge N_F\xi_{u,k}, u=1,2,\dots,U, \\
& k=1,2,\dots,K, \\
& \alpha_{u,n}+2\ln(|{\boldsymbol{\omega}_{u,n}^k}^{\text{H}}{\boldsymbol\gamma}_{u,n}|)- \\ &\ln(\sum_{l=1}^{U}e^{\alpha_{l,n}}
|{\boldsymbol{\omega}_{u,n}^k}^{\text{H}}{\boldsymbol\gamma}_{l,n}|^2\bar{\Delta}_{k}+
\sigma^2||{\boldsymbol{\omega}_{u,n}^k}||^2)\ge \\ & Y(t_{u,n}^k,\hat{t}_{u,n}^k), u=1,2,\dots,U, \\
& k=1,2,\dots,K, n=1,2,\dots,N_F, \\
& \ln\Big(\sum_{l=1}^{N_F}(1+\frac{2d^u_l}{N_F})e^{\alpha_{u,l}}+\\
& \frac{2}{N_F}\sum_{\substack{n,i=1\\i>n}}^{N_F}
{\eta_{nim}^+}^u
e^{\frac{1}{2}(\alpha_{u,n}+\alpha_{u,i})}\Big)\le T_m(\boldsymbol{\alpha}^u,\hat{\boldsymbol{\alpha}}^u), \\
& u=1,2,\dots,U, m=1,2,\dots,N_F,
\end{array}
\end{equation}
where ${\boldsymbol \alpha}^u=\{\alpha_{u,n}: n=1,2,\dots,N_F\}$. Now, the SCA algorithm can be used for problem \eqref{eq: local_conv_appr2} to find a local solution of the original problem. The complete algorithm is shown in \textbf{Algorithm} \ref{alg: SCA2}, where the superscript $^*$ denotes the optimal solution of \eqref{eq: local_conv_appr2}. Due to the concavity of the logarithm function, the linear approximation is always above the original function\footnote{By projecting the optimal solution from the approximated problem \eqref{eq: local_conv_appr2} to the original concave function (RHS in \eqref{eq: SCAVC constaint}) the constraint becomes loose and thus, the objective can always be reduced.\vspace{-0.3cm}}. Hence, \textbf{Algorithm} \ref{alg: SCA2} is guaranteed to monotonically converge to a local solution.

\begin{algorithm}
\caption{Successive convex approximation algorithm.}
\label{alg: SCA2}
\begin{minipage}{\columnwidth}
\begin{algorithmic}[1]
\STATE Set $\hat{t}_{u,n}^k=\hat{t}_{u,n}^{k(0)}, \forall u,k,n$ and $\hat{\boldsymbol{\alpha}}_{u,n}=\hat{\boldsymbol{\alpha}}_{u,n}^{(0)}, \forall u,n$.
\REPEAT
\STATE Solve Eq.\ \eqref{eq: local_conv_appr2}.
\STATE Update $\hat{t}_{u,n}^k={t}_{u,n}^{k(*)}, \forall u,k,n$ and $\hat{\boldsymbol{\alpha}}_{u,n}=\hat{\boldsymbol{\alpha}}_{u,n}^{(*)}, \forall u,n$.
\UNTIL Convergence.
\end{algorithmic}
\end{minipage}
\end{algorithm}
\vspace{-0.2cm}
\section{Numerical Results} \label{sec: simulation results}
\vspace{-0.1cm}
In this section, numerical results will be shown to demonstrate the performance of the proposed algorithm. SCAs presented in previous sections were derived for fixed receiver. The joint optimum can be achieved via alternating optimization \cite{Tervo-Tolli-Karjalainen-Matsumoto-13} which means that the problem is split to the optimization of transmit power for fixed receiver and optimization of receiver for fixed power allocation. Alternating between these two optimization steps converges to a local solution.

\begin{figure}[tbp!] \vspace{-0.2cm}
\centering
\includegraphics[angle=-90,width=\columnwidth]
{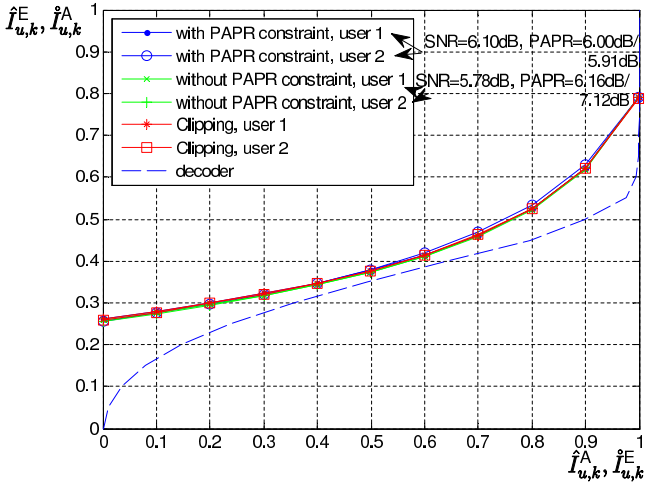}
\caption{EXIT chart for turbo equalizer with 6 dB PAPR threshold. $U=2$, $N_F=8$, $N_R=2$, $\hat{I}_u^{\text{E,target}}=0.7892$, $u=1,2$, $\mathring{I}_u^{\text{E,target}}=0.9998$, $\forall u$, $\epsilon_u=0.01$, $\forall u$, $N_L=5$.}
\label{fig: EXIT for 6dB and 1e-5} \vspace{-0.2cm}
\end{figure}

\begin{figure}[tbp!] \vspace{-0.3cm}
\centering
\includegraphics[angle=-90,width=\columnwidth]
{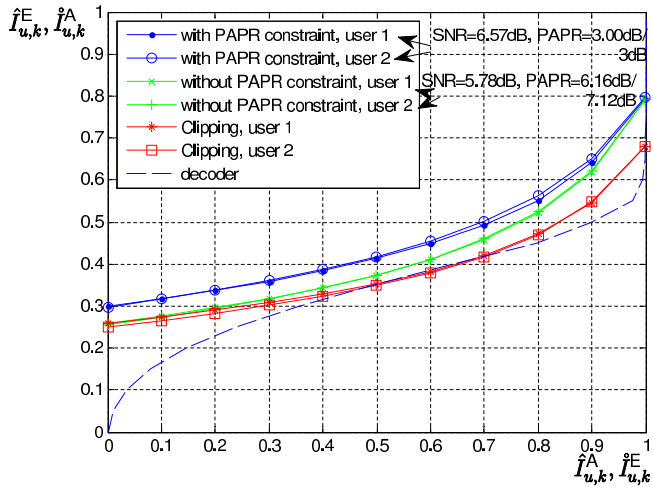}
\caption{EXIT chart for turbo equalizer with 3 dB PAPR threshold. $U=2$, $N_F=8$, $N_R=2$, $\hat{I}_u^{\text{E,target}}=0.7892$, $u=1,2$, $\mathring{I}_u^{\text{E,target}}=0.9998$, $\forall u$, $\epsilon_u=0.01$, $\forall u$, $N_L=5$.}
\label{fig: EXIT for 3dB and 1e-5} \vspace{-0.7cm}
\end{figure}

The following parameters is used in simulations: $U=2$, $N_R=2$, $N_F=8$, QPSK with Gray mapping, and systematic repeat accumulate (RA) code \cite{Divsalar-Jin-McEliece-98} with a code rate 1/3 and 8 internal iterations are used. The signal-to-noise ratio per receiver antenna averaged over frequency bins is defined by SNR$=\text{tr}\{\mathbf{P}\}/(N_RN_F\sigma^2)$. The channel we consider is a quasi-static Rayleigh fading 5-path average equal gain channel. The EXIT curve of the decoder is obtained by using 200 blocks for each a priori value with the size of a block being 6000 bits. The EXIT curves for the equalizer shown in Figs.\ \ref{fig: EXIT for 6dB and 1e-5} and \ref{fig: EXIT for 3dB and 1e-5} are obtained by averaging over 200 channel realizations. We will consider three different transmission strategies: power allocation with PAPR constraint, i.e., \textbf{Algorithm} \ref{alg: SCA2}, CCPA without PAPR constraint, i.e., \textbf{Algorithm} \ref{alg: SCA}, and amplitude clipping \cite{Gacanin-Takaoka-Adachi-06} applied to CCPA precoded transmission.

EXIT chart for the system with PAPR threshold being 6 dB and the MI targets being $(\mathring{I}_u^{\text{E,target}},\hat{I}_u^{\text{E,target}})=(0.9998,0.7892)$, $u=1,2$ is depicted in Fig.\ \ref{fig: EXIT for 6dB and 1e-5}. MI target can be converted to bit error probability (BEP) by using the equation \cite{tenBrink-01} \vspace{-0.1cm}
\begin{equation}
P_b\approx\frac{1}{2}\text{erfc}\Bigg(\frac{\sqrt{\text{J}^{-1}(\hat{I}_1^{\text{A,target}})
+\text{J}^{-1}(\hat{I}_1^{\text{E,target}})
}}{2\sqrt{2}}\Bigg). \vspace{-0.15cm}
\end{equation}
Hence, $(\mathring{I}_u^{\text{E,target}},\hat{I}_u^{\text{E,target}})=(0.9998,0.7892)$ corresponds to BEP $10^{-5}$.
It can be seen from Fig.\ \ref{fig: EXIT for 6dB and 1e-5} that there is not much difference between the PAPR constrained result and the one without PAPR constraint when the threshold is 6 dB. Furthermore, clipping the signal when the power is higher than 6 dB from the average power do not have significant impact on the results. The convergence point for algorithms with and without the PAPR constraint is indeed the preset target point. PAPRs without PAPR constraint are 6.16 dB and 7.12 dB for user 1 and user 2, respectively. PAPRs with PAPR constraint are at most 6 dB for both users. However, with PAPR constraint the SNR required to achieve the target point is 0.32 dB larger. After clipping the convergence points are (0.9998,0.7892) and (0.9998,0.7868) corresponding the BEPs $10^{-5}$ and $1.01\cdot10^{-5}$ for user 1 and user 2, respectively.

EXIT chart for the system with PAPR threshold being 3 dB and the MI targets being $(\mathring{I}_u^{\text{E,target}},\hat{I}_u^{\text{E,target}})=(0.9998,0.7892)$, $u=1,2$ is depicted in Fig.\ \ref{fig: EXIT for 3dB and 1e-5}. Now, we can see the impact of PAPR constraint which causes 0.79 dB increase of required SNR. However, the PAPR never exceeds 3 dB and the convergence point is still guaranteed to be the preset target point. The EXIT curves for clipping intersect the decoder curve at a low MI value, and the convergence points are (0.5142,0.3576) and (0.4629,0.3393) corresponding to BEPs 0.0933 and 0.1072. This was expected due to the fact that amplitude clipping causes distortion and hence, reduces the SNR and therefore MI after detection.

CCPA performs the power allocation such that the gap between the EXIT curves is larger than or equal to $\epsilon_u$. If we decrease $\epsilon_u$, the power consumption is reduced while the number of iterations in the equalizer increases \cite{Tervo-Tolli-Karjalainen-Matsumoto-13}. If clipping is used and $\epsilon_u$ is small, the EXIT curves of the equalizer and the decoder may intersect already at very low MI point which results in very high BEP. Therefore, PAPR constraint is crucial when CCPA is used with small $\epsilon_u$.
\vspace{-0.1cm}
\section{Conclusions} \label{Sec: conclusions}
\vspace{-0.1cm}
In this paper, we have derived the peak-to-average power ratio (PAPR) constrained power allocation problem for iterative FD-SC-MMSE multiuser SIMO detector. We derived an analytical expression of PAPR as a function of transmit power allocation and QPSK modulated symbol sequence. Moreover, a successive convex approximation for PAPR constrained problem was derived. Numerical results indicate that PAPR constraint is of crucial importance to guarantee the convergence of the iterative equalizer. The constraint derived in this paper is especially beneficial for the users on the cell edge due to the power limited transmission.

In this paper, we have presented our first results considering PAPR constrained power allocation and the aim of this paper is to provide more insight into the problem. This type of power loading requires centralized design, i.e., the base station reports the power allocations to each user. Development towards distributed solution is left as future work.
\vspace{-0.2cm}
\bibliographystyle{IEEEtran}
\bibliography{jour_short,conf_short,CCPA_MUSIMObib}

\begin{thebibliography}{10}
\providecommand{\url}[1]{#1}
\csname url@samestyle\endcsname
\providecommand{\newblock}{\relax}
\providecommand{\bibinfo}[2]{#2}
\providecommand{\BIBentrySTDinterwordspacing}{\spaceskip=0pt\relax}
\providecommand{\BIBentryALTinterwordstretchfactor}{4}
\providecommand{\BIBentryALTinterwordspacing}{\spaceskip=\fontdimen2\font plus
\BIBentryALTinterwordstretchfactor\fontdimen3\font minus
  \fontdimen4\font\relax}
\providecommand{\BIBforeignlanguage}[2]{{%
\expandafter\ifx\csname l@#1\endcsname\relax
\typeout{** WARNING: IEEEtran.bst: No hyphenation pattern has been}%
\typeout{** loaded for the language `#1'. Using the pattern for}%
\typeout{** the default language instead.}%
\else
\language=\csname l@#1\endcsname
\fi
#2}}
\providecommand{\BIBdecl}{\relax}
\BIBdecl

\bibitem{Pancaldi-Vitetta-Kalbasi-Al-Dhahir-Uysal-Mheidat-08}
F.~Pancaldi, G.~Vitetta, R.~Kalbasi, N.~Al-Dhahir, M.~Uysal, and H.~Mheidat,
  ``Single-carrier frequency domain equalization,'' \emph{Signal Processing
  Magazine, IEEE}, vol.~25, no.~5, pp. 37--56, 2008.

\bibitem{Slimane-07}
S.~Slimane, ``Reducing the peak-to-average power ratio of ofdm signals through
  precoding,'' \emph{Vehicular Technology, IEEE Transactions on}, vol.~56,
  no.~2, pp. 686--695, 2007.

\bibitem{Falconer-11}
D.~Falconer, ``Linear precoding of ofdma signals to minimize their
  instantaneous power variance,'' \emph{Communications, IEEE Transactions on},
  vol.~59, no.~4, pp. 1154--1162, 2011.

\bibitem{Yuen-Farhang-Boroujeny-12}
C.~Yuen and B.~Farhang-Boroujeny, ``Analysis of the optimum precoder in
  sc-fdma,'' \emph{Wireless Communications, IEEE Transactions on}, vol.~11,
  no.~11, pp. 4096--4107, 2012.

\bibitem{Yuan-Guo-Wang-Ping-08}
X.~Yuan, Q.~Guo, X.~Wang, and L.~Ping, ``Evolution analysis of low-cost
  iterative equalization in coded linear systems with cyclic prefix,''
  \emph{{IEEE} J. Select. Areas Commun.}, vol.~26, no.~2, pp. 301--310, Feb.
  2008.

\bibitem{Karjalainen-Codreanu-Tolli-Juntti-Matsumoto-11}
J.~Karjalainen, M.~Codreanu, A.~T\"{o}lli, M.~Juntti, and T.~Matsumoto,
  ``{EXIT} chart-based power allocation for iterative frequency domain {MIMO}
  detector,'' \emph{{IEEE} Trans. Signal Processing}, vol.~59, no.~4, pp.
  1624--1641, Apr. 2011.

\bibitem{tenBrink-01}
S.~ten Brink, ``Convergence behavior of iteratively decoded parallel
  concatenated codes,'' \emph{{IEEE} Trans. Commun.}, vol.~49, no.~10, pp.
  1727--1737, Oct. 2001.

\bibitem{Tervo-Tolli-Karjalainen-Matsumoto-12}
V.~Tervo, A.~T\"{o}lli, J.~Karjalainen, and T.~Matsumoto, ``On convergence
  constraint precoder design for iterative frequency domain multiuser {SISO}
  detector,'' in \emph{Proc. Annual Asilomar Conf. Signals, Syst., Comp.},
  Pacific Grove, CA, USA, Nov.4--7 2012, pp. 473--477.

\bibitem{Tervo-Tolli-Karjalainen-Matsumoto-13}
------, ``Convergence constrained multiuser transmitter-receiver optimization
  in single carrier {FDMA},'' \emph{{IEEE} Trans. Signal Processing}, 2013,
  (under review).

\bibitem{Brannstrom-Rasmussen-Grant-05}
F.~Br\"{a}nnstr\"{o}m, L.~K. Rasmussen, and A.~J. Grant, ``Convergence analysis
  and optimal scheduling for multiple concatenated codes,'' \emph{{IEEE} Trans.
  Inform. Theory}, vol.~51, no.~9, pp. 3354--3364, Sep. 2005.

\bibitem{Karjalainen-11}
\BIBentryALTinterwordspacing
J.~Karjalainen, ``Broadband single carrier multi-antenna communications with
  frequency domain turbo equalization,'' Ph.D. dissertation, University of
  Oulu, Oulu, Finland, 2011. [Online]. Available:
  \url{http://herkules.oulu.fi/isbn9789514295027/isbn9789514295027.pdf}
\BIBentrySTDinterwordspacing

\bibitem{Boyd-Vandenberghe-04}
S.~Boyd and L.~Vandenberghe, \emph{Convex Optimization}.\hskip 1em plus 0.5em
  minus 0.4em\relax Cambridge, U.K.: Cambridge Univ. Press, 2004.

\bibitem{Divsalar-Jin-McEliece-98}
D.~Divsalar, H.~Jin, and R.~J. McEliece, ``Coding theorems for 'turbo-like'
  codes,'' in \emph{Proc. Annual Allerton Conf. Commun., Contr., Computing},
  Urbana, Illinois, USA, Sep.23--25 1998, pp. 201--210.

\bibitem{Gacanin-Takaoka-Adachi-06}
H.~Gacanin, S.~Takaoka, and F.~Adachi, ``Reduction of amplitude clipping level
  with {OFDM/TDM},'' in \emph{Vehicular Technology Conference, 2006. VTC-2006
  Fall. 2006 IEEE 64th}, 2006, pp. 1--5.

\end{thebibliography}

\end{document}